\documentclass[reprint,twocolumn,aps,
showkeys,
notitlepage]{revtex4-2}

\usepackage{
amsmath,
amssymb,
comment,
}
\usepackage{bm}
\usepackage{color}  
\usepackage{extarrows}
\usepackage{appendix}
\usepackage{paralist}

\usepackage[hyperindex=true,
          pdfstartview=FitH,
          bookmarksnumbered=true,
          bookmarksopen=true,
          citecolor=blue,
          linkcolor=blue,
          colorlinks=true,
          pdfborder=001,
          unicode]{hyperref}

\allowdisplaybreaks

\newcommand{\D}{{\rm d}}
\newcommand{\ie}{\emph{i.e. }}
\newcommand{\eg}{\emph{e.g. }}

\renewcommand\({\left(}
\renewcommand\){\right)}

\DeclareMathOperator{\Tr}{Tr}

\newcommand{\blue}[1]{\textcolor{blue}{#1}}
\newcommand{\red}[1]{\textcolor{red}{#1}}
\renewcommand{\blue}[1]{\textcolor{black}{#1}}
\renewcommand{\red}[1]{\textcolor{black}{#1}}

\newcommand{\ind}[2]{^{#1}{}_{#2}}

\newcommand{\re}{{\rm e}}
\newcommand{\pfrac}[2]{\frac{\partial #1}{\partial #2}}
\newcommand{\mean}[1]{\big\langle #1\big\rangle}

\begin{document}

\title{Fluctuation Theorem on a Riemannian Manifold}

\author{Yifan Cai}
\email{caiyifan@mail.nankai.edu.cn}
\affiliation{School of Physics, Nankai University, Tianjin 300071, China}

\author{Tao Wang}
\email{taowang@mail.nankai.edu.cn}
\affiliation{School of Physics, Nankai University, Tianjin 300071, China}

\author{Liu Zhao}
\email{lzhao@nankai.edu.cn}
\affiliation{School of Physics, Nankai University, Tianjin 300071, China}

\date{January 1, 2024}

\begin{abstract}
Based on the covariant underdamped and overdamped Langevin equations 
with Stratonovich coupling to multiplicative noises and the associated 
Fokker-Planck equations on Riemannian manifold, we present  
the first law of stochastic thermodynamics on the trajectory level. 
The corresponding fluctuation theorems are also
established, with the total entropy production of the Brownian particle 
and the heat reservoir playing the role of dissipation function.

\end{abstract}

\maketitle
\section{Introduction}

Since the early 1990s, fluctuation theorems (FTs) \cite{evans1994equilibrium,
gallavotti1995dynamical,crooks1999entropy} have played an 
indispensable role in understanding the origin of 
macroscopic irreversibility. Such theorems, often realized in the 
form of unequal probabilities for the forward and reversed processes, greatly helped in 
resolving the long lasting puzzles and debates regarding Boltzmann's H-theorem, 
known as Loschmidt paradox. Based on Sekimoto's work \cite{sekimoto1998langevin} 
on stochastic energetics, Seifert \cite{seifert2005entropy} was able to establish a 
version of FT associated to the stochastic trajectories 
described by overdamped Langevin equation (OLE), and subsequent works 
\cite{imparato2006fluctuation,chernyak2006path,ohkuma2007fluctuation} extended 
the construction to the cases of various generalized forms of Langevin equation (LE). 

{\em Can we establish FTs on the trajectory level
on curved Riemannian manifold?} This is the question we wish to address in this work. 
In recent years, stochastic thermodynamics has gained increasing importance 
in understanding phenomena at the mesoscopic scale \cite{randel2004molecular, 
muthukumar2006simulation, seifert2012stochastic, qian2016entropy, 
golokolenov2023thermodynamics}. There are certain realistic scenarios,
such as the diffusion of individual protein molecule on a biological membrane, 
which calls for a construction of LE and FT 
on Riemannian manifolds. Another motivation for the quest of stochastic thermodynamics 
and FT on Riemannian manifolds is to take it as a midway step towards 
general relativistic description of these fields, which is important because, 
in essence, every physics system must abide by the principles of relativity, 
whereas the spacetime symmetries in general relativity 
impose much stronger restrictions, making
it harder to break the time reversal symmetry.

\blue{Historically,} there have been some
repeated attempts in the construction of LE on Riemannian manifolds 
in the mathematical \cite{Elworthy1998, hsu2002stochastic, 10.1007/BFb0088730, 
emery2012stochastic, armstrong2016coordinate, armstrong2018intrinsic} 
\blue{and physical \cite{robert1977covariant, robert1985covariant}}
literature. However, the first law and other thermodynamic 
relations were not considered in these works. Xing \textit{et al.} \cite{ding2020covariant, 
ding2022covariant} explored an Ito-type nonlinear LE and established 
a FT on Riemannian manifold. \blue{Most of these works 
have interpreted the word ``covariance'' 
in the sense of second order (or {\em jet bundle}) geometry, which is 
different from the usual coordinate covariance in 
standard Riemannian geometry. Ref. \cite{robert1977covariant} seems to be an exception.
However, the difference between our work and \cite{robert1977covariant} is acute: 
the LE in \cite{robert1977covariant} is presented only in the configuration space, 
something like the OLE to be described in Section \ref{Overdamped-Langevin-dynamics}. 
However, as will be explained near the end of Section \ref{Overdamped-Langevin-dynamics}, 
the covariance of LE in \cite{robert1977covariant}
actually holds only in flat manifolds and only with respect to 
linear coordinate transformations, while our work covers both the configuration space (OLE)
and phase space (ULE) descriptions, and our formalism is genuinely covariant under general
coordinate transformation in the sense of first order geometry.}

\section{Langevin and Fokker-Planck equations on Riemannian manifold}
\label{Langevin-dynamics-on-Riemann-manifold}

\subsection{Underdamped case}
\label{Underdamped-Langevin-dynamics}

In Ref.~\cite{cai2023relativistic}, the relativistic 
covariant underdamped Langevin equation (ULE) on {\em pseudo-Riemannian spacetime} 
is established. The same procedure can be used for constructing the LE 
\blue{for a point particle of mass $m>0$ moving 
on $d$-dimensional} {\em Riemannian space} $M$ with positive definite 
metric $g_{\mu\nu}(x)$, so we directly present the result, 
\begin{align}
\D \tilde x_t^\mu&=\frac{\tilde p_t^\mu}{m}\D t,\label{la3.1}\\
\D\tilde p_t^\mu &=\left[R\ind{\mu}{a}\circ_S\D\tilde w^a_t+\frac{1}{2}R\ind{\mu}{a}
\frac{\partial}{\partial p^\nu}R\ind{\nu}{a}\D t\right]\nonumber\\
&\quad-\frac{1}{m}K\ind{\mu}{\nu}\tilde p_t^\nu\D t+f^\mu_{\text{ex}}\D t
-\frac{1}{m}\Gamma\ind{\mu}{\alpha\beta}
\tilde p_t^\alpha \tilde p_t^\beta\D t,\label{la3.2}
\end{align}
where $R\ind{\mu}{a}$ represent the stochastic amplitudes 
which may depend on $\tilde x^\mu$ and $\tilde p^\mu$, \blue{$\Gamma\ind{\mu}{\alpha\beta}$ is the 
Christoffel connection associated with $g_{\mu\nu}(x)$, $K\ind{\mu}{\nu}$ is the 
tensorial damping coefficient (referred to as damping tensor henceforth)}, 
and $\D\tilde w^a_t$ are Gaussian noises with probability distribution functions (PDFs)
\begin{align}\label{wienner}
\Pr[\D\tilde w_t^a=\D w^a]
=\frac{1}{(2\pi \D t )^{d/2}}
\exp\left[-\frac{\delta_{ab}\D w^a \D w^b}{2\D t} \right].
\end{align}
The symbol $\circ_S$ represents the Stratonovich coupling which ensures the chain rule 
in stochastic calculus. 
\red{Greek indices} \blue{$\mu,\nu,\cdots$} 
\red{label spatial directions and latin indices} \blue{$a,b,\cdots$} 
\red{label independent noises, 
which are all running from 1 to $d$.} 
Tilded variables like \blue{$\tilde x^\mu, \tilde p^\mu$}
represent random variables, and the un-tilded symbols like \blue{$x^\mu, p^\mu$} 
their realizations. \blue{In particular, $(x^\mu, p^\mu)$ denotes the coordinate of 
the Brownian particle on $TM$ in a concrete realization.}

Some remarks are in due here.

(i) The LE described above is a system of stochastic differential equations  
on the tangent bundle $TM$ 
regarded as the space of micro states of the Brownian particle. \blue{Normally, the 
space of micro states for a particle is taken to be the cotangent bundle $T^\ast M$. 
However, due to the non-degeneracy of the metric $g_{\mu\nu}$, the tangent and cotangent 
spaces are dual to each other, and the tangent and cotangent bundles can be 
used interchangeably (see \cite{sarbach2013rel, sarbach2014geo,Acuna-Cardenas} for 
explicit use of both approaches). }
Recall that \red{$TM$ is naturally equipped with the Sasaki metric 
\cite{sarbach2013rel, sarbach2014geo,sasaki1958differential}}
\begin{align*}
\hat g:=g_{\mu\nu}\D x^\mu\otimes\D x^\nu+g_{\mu\nu}\theta^\mu\otimes\theta^\nu,
\end{align*} 
where $\theta^\mu=\D p^\mu+\Gamma\ind{\mu}{\alpha\beta}p^\alpha\D x^\beta$, 
together with the invariant volume element [here $g(x):=\det g_{\mu\nu}(x)$]
\begin{align*}
\D^{2d} X&=g(x)~ \D x^1\wedge\D x^2\wedge...\wedge
\D x^{d}\wedge\D p^1\wedge...\wedge\D p^{d}.
\end{align*}
\blue{As explained in \cite{sarbach2013rel, sarbach2014geo}, 
the Sasaki metric is closely connected to the symplectic structure on $TM$. }

\blue{Notice that the metric $g_{\mu\nu}(x)$ on $M$ plays an indispensable role while
obtaining the above Sasaki metric and also while describing the last term of 
eq.~\eqref{la3.2}.}

\blue{(ii) The choice for the stochastic amplitude $R\ind{\mu}{a}$ is non-unique. 
Different choices correspond to different Langevin systems, and the result of this 
work should hold for all choices such that 
$R\ind{\mu}{a}$ is an invertible matrix function which is differentiable in $(x^\mu,p^\mu)$  
and transforms as a vector for each fixed $a$.}

(iii) Our approach to LE is the {\em traditional} one, as opposed to the more abstract 
nonlinear approach used in \cite{ding2020covariant, ding2022covariant}.
Eq.~\eqref{la3.2} can be viewed as 
the geodesic equation supplemented by additional force terms, including a 
stochastic force
\begin{align*}
\xi^\mu:=R\ind{\mu}{a}\circ_S\D\tilde w^a_t/\D t
+\frac{1}{2}R\ind{\mu}{a}\frac{\partial}{\partial p^\nu}R\ind{\nu}{a},
\end{align*}
a damping force $ f^\mu_{\text{dp}}:=-K\ind{\mu}{\nu}\tilde p_t^\nu/m$ and an 
external force $f^\mu_{\text{ex}}$. The second term in the stochastic force 
is known as {\em additional stochastic force} \cite{klimontovich1994nonlinear,
cai2023relativistic2}, which is required in order for the Brownian particle to be able to
reach thermal equilibrium with the heat reservoir. 
In one-dimensional case, 
the stochastic force can also be expressed in the form of post-point rule 
$\xi=R\circ_p\D\tilde w_t/\D t$, hence some authors argued that the post-point rule 
is better suited for LE with multiplicative noises. However, in higher-dimensional 
cases, the post-point rule leads to a different result,  
\[
R\ind{\mu}{a}\circ_p\D\tilde w^a_t/\D t
=R\ind{\mu}{a}\circ_S\D\tilde w^a_t/\D t
+\frac{1}{2}R\ind{\nu}{a}\pfrac{}{p^\nu}R\ind{\mu}{a}.
\]

(iv) The Stratonovich coupling \blue{maintains the chain rule, which} 
ensures the covariance in the usual sense in Riemannian geometry. 
This \blue{makes an important difference} from the previous works 
\cite{10.1007/BFb0088730,emery2012stochastic,armstrong2016coordinate,armstrong2018intrinsic,
ding2020covariant, ding2022covariant}.

(v) Although eqs.~\eqref{la3.1}-\eqref{la3.2} look the same as their 
relativistic counterparts \cite{cai2023relativistic}, there are some essential
differences. First, the time $t$ 
used here is absolute, meaning that eqs.~\eqref{la3.1}-\eqref{la3.2} are 
{\em non-relativistic}; Second, the heat reservoir hiding behind the stochastic 
and damping force terms is also non-relativistic, there is no need to worry about the 
relativistic effects such as the Tolman-Ehrenfest red shift;  
Last, the momentum space is flat, as opposed to the relativistic case.

The external force term depends on the position $x$ of the Brownian particle 
and an external control parameter $\lambda$, and can be separated into  
conservative and non-conservative parts, 
\begin{align*}
f^\mu_{\text{ex}}=f^\mu_{\text{con}}+f^\mu_{\text{noc}}
=-\nabla^\mu U(x,\lambda)+f^\mu_{\text{noc}}.
\end{align*}
There is some ambiguity in this decomposition, \eg
\begin{align*}
f^\mu_{\text{ex}}=(f^\mu_{\text{con}}+\hat f^\mu)
+(f^\mu_{\text{noc}}-\hat f^\mu)
=\hat{f}^\mu_{\text{con}}+\hat f^\mu_{\text{noc}},
\end{align*}
where $\hat f^\mu$ is an arbitrary conservative force. 
This also leads to an ambiguity in the energy  
\begin{align*}
E(x,p,\lambda)=\frac{1}{2m}g_{\mu\nu}(x)p^\mu p^\nu+U(x,\lambda)
\end{align*}
of the Brownian particle. In the extremal case with $\hat f^\mu=-f^\mu_{\text{con}}$, 
The energy will be consisted purely of the kinematic energy. 

Since the Stratonovich coupling preserves the chain rule, we have
\begin{align}
\D \tilde E_t&=\pfrac{E}{p^\mu}\D\tilde p^\mu_t
+\pfrac{E}{x^\mu}\D\tilde x_t^\mu+\D_\lambda U \notag\\
&=\frac{1}{m}(\tilde p_t)_\mu (\xi^\mu+f^\mu_{\text{dp}}) \D t
+\frac{1}{m}(\tilde p_t)_\mu f^\mu_{\text{noc}}\D t+\D_\lambda U,
\label{DEt}
\end{align}
where $\tilde E_t$ is the energy considered as a random variable and $E$ its realization.
The part of the increase of energy caused by the heat reservoir is \blue{purely a thermal effect
and thus} comprehended as {\em trajectory heat} $\D \tilde Q_t$, 
the rest part \blue{is purely mechanical and should be comprehended}
as {\em trajectory work} $\D \tilde W_t$, 
\begin{align}\label{underdamped-heat-work}
&\D \tilde Q_t=\frac{(\tilde p_t)_\mu}{m}\(\xi^\mu+f^\mu_{\text{dp}}\)\D t,\nonumber\\
&\D \tilde W_t=\frac{(\tilde p_t)_\mu}{m} f^\mu_{\text{noc}}\D t+\D_\lambda U.
\end{align}
Thus eq.~\eqref{DEt} becomes the first law of stochastic thermodynamics
on the trajectory level, \ie 
\[
\D \tilde E_t=\D \tilde Q_t+\D \tilde W_t.
\]
The ambiguity in the decomposition of the external force
also leads to an ambiguity in the trajectory work. However, 
in any case, the trajectory heat 
$\D \tilde Q_t=\D \tilde E_t-\D \tilde W_t$ is always unambiguously defined. 
\blue{Please be reminded that, unlike the usual heat and work in standard thermodynamics 
which are inexact differentials defined on the space of macro states, 
the {\em trajectory heat and work} are only defined 
on a stochastic trajectory of the Brownian particle.}

The Fokker-Planck equation (FPE) associated to
eqs.~\eqref{la3.1}-\eqref{la3.2} can also be established on $TM$.  
The PDF for the Brownian particle under the 
measure $\D^{2d} X$ is denoted as $\Phi_t(x,p):=\Pr[\tilde x_t=x,\tilde p_t=p]$, 
and is clearly coordinate-independent. By use of the diffusion operator method 
\cite{bakry2014analysis}, one can get
\begin{align}
\partial_t\Phi_t&=\pfrac{}{p^\mu}\left[\frac{1}{2}D^{\mu\nu}\pfrac{}{p^\nu}\Phi_t
+\frac{1}{m}K\ind{\mu}{\nu}p^\nu\Phi_t-f^\mu_{\text{ex}}\Phi_t \right]\nonumber\\
&\quad-\frac{1}{m}\mathcal{L}(\Phi_t),
\label{fokkerplank1}
\end{align}
where $D^{\mu\nu}:=R\ind{\mu}{a}R\ind{\nu}{a}$ is the diffusion tensor \red{and  
\[
\mathcal L=p^\mu\pfrac{}{x^\mu}
-\Gamma\ind{\mu}{\alpha\beta}p^\alpha p^\beta\pfrac{}{p^\mu}
\] 
is the Liouville vector field on $TM$ \cite{sarbach2013rel, sarbach2014geo}.} 

Let the non-conservative force be temporarily turned off and the external control 
parameter be fixed. Then, after sufficiently long period of time, 
the Brownian particle will reach a thermal equilibrium with the
heat reservoir, yielding the equilibrium PDF
\begin{align}
\Phi_t(x,p)=\frac{1}{Z}\exp\left[-\frac{1}{T}\left(\frac{g_{\mu\nu}p^\mu p^\nu}{2m}
+U(x,\lambda)\right)\right].
\label{equilibrium1}
\end{align}
Putting this PDF into the FPE \eqref{fokkerplank1}, one gets  
the Einstein relation
\begin{align}
D^{\mu\nu}=2TK^{\mu\nu},\label{Einstein1}
\end{align}
\blue{which implies that the damping tensor $K^{\mu\nu}$ are not independent of
the stochastic amplitudes $R\ind{\mu}{a}$ and that $K^{\mu\nu}$ is invertible as a matrix. 
As long as only the FPE is concerned, 
there is an additional freedom in the sign choice of $R\ind{\mu}{a}$, because $D^{\mu\nu}$
appear as a quadratic form in $R\ind{\mu}{a}$.} 

To facilitate the discussion about FT, we introduce 
the time-reversal transform (TRT) \blue{for the process ranging from $t_I$ to $t_F$:
\begin{align*}
I: \left\{
\begin{array}{l} x^\mu(t) \mapsto x^\mu(t_F + t_I -t)\cr
p^\mu(t)\mapsto -p^\mu(t_F + t_I -t)
\end{array}
\right.,
\end{align*}
which is often briefly described as $I:(x,p)\mapsto(x,-p)$ for short. Notice that 
the infinitesimal time increment $\D t$ is not affected by such transformation
and remains to be positive.
}

It is obvious that TRT preserves the metric, 
\ie $I^*\hat g=\hat g$. 
The damping force $f^\mu_{\text{dp}}|_X=-K\ind{\mu}{\nu}p^\nu/m$ reverses 
sign under TRT, thus the damping tensor must be invariant under TRT,
\begin{align*}
K^{\mu\nu}|_{I(X)}=K^{\mu\nu}|_X, \quad X=(x,p),\quad I(X) =(x,-p).
\end{align*}
Eq.~\eqref{Einstein1} implies that the diffusion tensor 
is also invariant under TRT. There is some freedom in choosing 
the stochastic amplitudes, and hence also in determining their behaviors under TRT.
Here we assume the simplest transformation rule,
\begin{align*}
R\ind{\mu}{a}|_{I(X)}=R\ind{\mu}{a}|_{X}.
\end{align*} 
An immediate consequence is that the additional stochastic force should 
reverse sign under TRT. The coefficients of $\D t$ in eq.~\eqref{la3.2}
can be classified into even and odd parts under TRT, \ie
\begin{align*}
F^\mu=f^\mu_{\text{ex}}-\frac{1}{m}\Gamma\ind{\mu}{\alpha\beta}p^\alpha p^\beta
\text{\,  with  \,}
F^\mu|_{I(X)}=F^\mu|_X,
\end{align*}
and
\begin{align*}
\bar{F}^\mu=\frac{1}{2}R\ind{\mu}{a}\pfrac{}{p^\nu}R\ind{\nu}{a}
-\frac{1}{m}K\ind{\mu}{\nu}p^\nu
\text{\,  with  \,}
\bar{F}^\mu|_{I(X)}=-\bar{F}^\mu|_X.
\end{align*}
Then eq.~\eqref{la3.2} can be recast in a simpler form,
\begin{align}\label{la3.3}
\D\tilde p^\mu_t=R\ind{\mu}{a}\circ_S\D\tilde w^\mu_t+F^\mu\D t+\bar{F}^\mu\D t.
\end{align}
\blue{The last equation is the starting point for introducing the discretized version of the
ULE in Appendix \ref{Continuity-limit-underdamped}.}

\subsection{Overdamped case}
\label{Overdamped-Langevin-dynamics}

The stochastic mechanics characterized by LE is a branch of 
physics with multiple time scales, of which the smallest one is the time scale 
$\D t$ which allows for a sufficient number of collisions between the Brownian particle
and the heat reservoir particles which cause little changes in the state of 
the Brownian particle \cite{sekimoto2010stochastic}. If the temporal resolution
$\Delta t$ greatly exceeds $\D t$ but is still smaller than the relaxation time, 
the ULE \eqref{la3.1}-\eqref{la3.2} emerges. 
If $\Delta t$ is further increased so that it is 
much larger than the relaxation time, the overdamped limit emerges. 

The relaxation process of the Brownian particle 
can be viewed either as the process by which the damping force attains 
a state of mechanical equilibrium with other forces (mechanical 
relaxation process), or as the process by 
which the Brownian particle achieves local thermal equilibrium with the 
heat reservoir (thermodynamic relaxation process). 
The characteristic timescales associated with both processes are 
$m/\kappa$, where $\kappa$ is the eigenvalue of $K^{\mu\nu}$. 
Consequently, the OLE should arise when $\Delta t$ greatly exceeds $m/\kappa$.

The two kinds of relaxation process correspond to two approaches 
for taking the overdamped limit. From the perspective of mechanical relaxation process, 
the OLE can be described as the condition for mechanical 
equilibrium
\begin{align}\label{overdampedLangevin}
0=f^\mu_{\text{dp}}+f^\mu_{\text{ex}}+\xi^\mu. 
\end{align}
This result is achieved in flat space under the condition that 
the damping tensor is position- and momentum-independent 
\cite{bodrova2016underdamped,huang2021quantitative}, 
and is often viewed as a stochastic differential equation in 
configuration space. However, if the stochastic amplitudes 
are momentum-dependent, so is the additional stochastic force. Thus the  
mechanical equilibrium condition cannot be understood 
as a stochastic differential equation in 
configuration space. To avoid the above difficulty, let us consider the 
simpler situation in which the stochastic amplitudes are momentum-independent.  
Using the thermodynamic relaxation approach \cite{durang2015overdamped}, 
it will be shown that, even in this simpler situation,  
a nontrivial additional stochastic force term still arises in the corresponding OLE.

The overdamped condition implies that the momentum space PDF already
reaches the equilibrium form, while the configuration space PDF does not, 
so that the full PDF $\Phi_t(x,p)$ can be factorized,
\begin{align*}
\Phi_t(x,p)=\rho_t(x)P^s(x,p),
\end{align*}
where 
\begin{align}\label{local-equilibrium}
P^s(x,p):=\frac{1}{(2\pi m T)^{-d/2}}\exp\left[-\frac{g_{\mu\nu}(x)p^\mu p^\nu}{2mT}\right]
\end{align}
is the (Maxwell) equilibrium PDF in momentum space, 
and $\displaystyle \rho_t(x):=\int g^{{1}/{2}}\D^d p ~\Phi_t(x,p)=\Pr[\tilde x_t=x]
$ is the PDF in configuration space. By adding the first order 
corrections from the near equilibrium states, the overdamped FPE is found to be
(see Appendix.~\ref{overdamped_limit}), 
\begin{align}\label{fokkerplank2}
\partial_t \rho_t=\nabla_\mu\left[\frac{1}{2}\hat D^{\mu\nu}\nabla_\nu \rho_t
-\hat K\ind{\mu}{\nu}f^{\nu}_{\text{ex}}\rho_t \right],
\end{align}
where $\hat K^{\mu\nu}=(K^{-1})^{\mu\nu}$ and $\hat D^{\mu\nu}=4T^2(D^{-1})^{\mu\nu}$. 
Two important properties of eq.~\eqref{fokkerplank2} are worth of notice: 
1) The Einstein relation still holds for the rescaled damping and diffusion tensors 
\begin{align*}
\hat D^{\mu\nu}=2T\hat K^{\mu\nu};
\end{align*}
and 2) The Boltzmann distribution
\begin{align*}
\rho_t(x)=\frac{1}{Z_x}\re^{-U(x,\lambda)/T}
\end{align*}
is a solution of eq.~\eqref{fokkerplank2} provided the non-conservative force is turned off 
and the external parameter is fixed, \blue{wherein $Z_x$ represents the configuration space
normalization factor, which should not be confused with the normalization 
factor $Z$ appeared in eq.~\eqref{equilibrium1}.}

Using the diffusion operator method, 
it can be checked that the LE associated with 
eq.~\eqref{fokkerplank2} takes the same form as eq.~\eqref{overdampedLangevin}, 
but with a momentum-independent stochastic force term
\begin{align}
\xi^\mu=R\ind{\mu}{a}\circ_S\D\tilde w^a_t/\D t
+\frac{1}{2}R\ind{\mu}{a}\nabla_\nu(\hat K\ind{\nu}{\alpha}R\ind{\alpha}{a}). 
\label{xinew}
\end{align}
Denoting
\begin{align}
\hat R\ind{\mu}{a}:=\hat K\ind{\mu}{\nu}R\ind{\nu}{a},\,
\hat F^\mu=\hat K\ind{\mu}{\nu}f^\nu_{\text{ex}}
+\frac{1}{2}\hat R\ind{\mu}{a}\nabla_\nu\hat R\ind{\nu}{a},
\label{stforce}
\end{align}
the OLE can be written in a simpler form
\begin{align}\label{la-overdamped-continuity}
\D\tilde x_t^\mu=\hat R\ind{\mu}{a}\circ_S\D\tilde w^a_t+\hat F^\mu\D t.
\end{align}
\blue{This equation is similar in form to the LE presented in \cite{robert1985covariant}. 
However, unlike eq.~\eqref{stforce}, the additional stochastic force 
presented in \cite{robert1985covariant} contains only 
an ordinary coordinate derivative rather than covariant derivative. 
Consequently, the claimed general covariance of the LE of \cite{robert1977covariant} 
is questionable: it actually holds only for 
flat manifolds and only with respects to linear coordinate transformations.}

Since the inertial effect can be ignored in the overdamped case, the energy of 
the Brownian particle contains only the potential energy, \ie $E(x,\lambda)=U(x,\lambda)$. 
Using eq.~\eqref{overdampedLangevin} and the chain rule, we have
\begin{align*}
\D\tilde E_t&=\pfrac{U}{x^\mu}\D\tilde x_t^\mu+\D_\lambda U
=-(f_{\text{con}})_\mu\D\tilde x_t^\mu+\D_\lambda U\notag\\
&=[\xi_\mu+(f_{\text{dp}})_\mu]\D\tilde x_t^\mu
+(f_{\text{noc}})_\mu\D\tilde x^\mu_t+\D_\lambda U.
\end{align*}
Similar to the underdamped case, the energy absorbed from the heat reservoir 
is understood as the trajectory heat, and the  
rest part of the energy increase as trajectory work. Thus we have,
thanks to eq.~\eqref{overdampedLangevin},
\begin{align}
&\D\tilde Q_t=[\xi_\mu+(f_{\text{dp}})_\mu]\D\tilde x_t^\mu
=-(f_{\text{ex}})_\mu\D\tilde x^\mu_t,\label{trajectory-heat}\\
&\D\tilde W_t=(f_{\text{noc}})_\mu\D\tilde x^\mu_t+\D_\lambda U.
\nonumber
\end{align}

\section{Fluctuation theorem}
\label{Fluctuation-theorem}

Now we come to the stage for describing FT on Riemannian manifold
based on the description of stochastic trajectories. 
Since the trajectory probability in continuous time is hard to deal with, 
we \blue{adopt the following strategy:} first we take a discrete equidistant 
set of time nodes $t_I=t_0<t_1<...<t_n=t_F$ \blue{to rewrite the LE and 
factorize the corresponding trajectory probabilities},
and then take the continuum limit at the end of the calculation. 

The stochastic process in discrete time can be viewed as a sequence of 
random variables, \ie $\tilde X_{[t]}=(\tilde X_0,\tilde X_1,...,\tilde X_n)$ 
with $\tilde X_i=\tilde X_{t_i}$ for the underdamped case and
$\tilde x_{[t]}=(\tilde x_0,\tilde x_1,...,\tilde x_n)$, 
with $\tilde x_i=\tilde x_{t_i}$ for the overdamped case. 

Before proceeding, it is necessary to clarify the 
concepts of {\em ensemble} and {\em trajectory} entropy productions. 
At the time $t$, the ensemble entropy of the Brownian particle reads 
\[ S_{t}=-\int \D^{2d} X \Phi_t\ln\Phi_t, 
\]
and the ensemble entropy production in the process is 
$\Delta S= S_{t_F}-S_{t_I}$. 
Notice that \blue{(throughout this paper, an overline denotes ensemble average, 
while $\langle~~\rangle$ denotes trajectory average)}
\[
-\overline{\ln\rho_t}:=
-\int g^{{1}/{2}}\D^d x~ \rho_{t} \ln\rho_t
\]
is {\em not} the entropy of the overdamped Brownian particle, however, the ensemble 
entropy production of the Brownian particle can be represented as the difference 
of $-\overline{\ln\rho_t}$, because the ensemble entropy for the overdamped Brownian particle 
can be evaluated to be
\begin{align*}
&S_t=-\overline{\ln\rho_t} +\frac{d}{2}+\frac{d}{2}\ln(2\pi mT), 
\end{align*}
\blue{where the last two terms arise from the momentum
space integration of the term involving the distribution $P^s(x,p)$ given in 
eq.~\eqref{local-equilibrium}. }
Subtracting the initial value from the final value leaves only the difference of
$-\overline{\ln\rho_t}$,
\begin{align*}
\Delta S=\overline{\ln\rho_{t_I}}-\overline{\ln\rho_{t_F}}.
\end{align*}
In contrast, the trajectory entropy production is defined simply 
to be the difference between the logarithms 
of the PDF at the initial and final times, \ie
\[ 
\Delta S_{X_{[t]}}=\ln\Phi_{t_I}(X_0)-\ln\Phi_{t_F}(X_n) 
\]
for the underdamped and
\[
\Delta S_{x_{[t]}}=\ln\rho_{t_I}(X_0)-\ln\rho_{t_F}(X_n) 
\]
for the overdamped cases.

Now let us consider the underdamped case. It is important to distinguish the 
terms {\em process} and {\em trajectory}: the latter is a realization of the former. 
The forward process $\tilde X_{[t]}$ refers to a stochastic process governed by the 
ULE \eqref{la3.1}-\eqref{la3.2}, wherein 
the external control parameter $\lambda_{[t]}=(\lambda_0,\lambda_1,...,\lambda_n)$ 
varies over time. Correspondingly, the reversed process $\tilde X^-_{[t]}$ also 
refers to a stochastic process governed by the same LE, 
but its initial state should be identified with the time-reversal of 
the final state of the forward process, \ie $\tilde X^-_0=I(\tilde X_n)$,
and the corresponding external control parameter should satisfy 
$\lambda^-_i=\lambda_{n-i}$. The reversed trajectory $X_{[t]}=(X_0, X_1,..., X_n)$ 
is defined such that $X_i^-:=I(X_{n-i})$. 

We will prove that the total entropy production, \ie the sum of the
trajectory entropy production with the change of the entropy of the 
heat reservoir, should be
\begin{align}
\Sigma_{X_{[t]}}=\ln\frac{\Pr[\tilde X_{[t]}= X_{[t]}]}{\Pr[\tilde X^-_{[t]}=X^-_{[t]}]}.
\label{Pfb}
\end{align}

Since the Brownian motion is a Markov process, the trajectory probability 
can be decomposed into product of transition probabilities,
\begin{align*}
&\Pr[\tilde X_{[t]}= X_{[t]}]\\
&\quad= \blue{\Big{(}}\prod_{i=0}^{n-1} 
\Pr[\tilde X_{i+1}=X_{i+1}|\tilde X_i=X_i]\blue{\Big{)}} \Pr[\tilde X_0= X_0].
\end{align*}
\begin{widetext}
A similar decomposition can be made for $\Pr[\tilde X^-_{[t]}=X^-_{[t]}]$. 
Therefore, we have
\begin{align}\label{SigmaXt}
\Sigma_{X_{[t]}}
&=\sum_{i=0}^{n-1}\ln\frac{\Pr[\tilde X_{i+1}=X_{i+1}|\tilde X_i=X_i]}
{\Pr[\tilde X_{i+1}^-=X_{i+1}^-|\tilde X_i^-=X_i^-]}
+\ln\frac{\Pr[\tilde X_0= X_0]}{\Pr[\tilde X^-_0= X^-_0]}\notag\\
&=\sum_{i=0}^{n-1}\ln\frac{\Pr[\tilde X_{i+1}=X_{i+1}|\tilde X_i=X_i]}
{\Pr[\tilde X_{n-i}^-=I(X_{i})|\tilde X_{n-i-1}^-=I(X_{i+1})]}
+\ln\frac{\Pr[\tilde X_0= X_0]}{\Pr[ \tilde X_n= X_n]},
\end{align}
\end{widetext}
where, in the last step, the definitions of the reversed process and 
reversed trajectory have been used and rearrange of terms in the summation 
has been adopted. 
The last term in eq.~\eqref{SigmaXt} is simply the trajectory 
entropy production, because 
\[
\Phi_{t_I}(X_0) = \Pr[\tilde X_0= X_0],\quad 
\Phi_{t_F}(X_n) = \Pr[\tilde X_n= X_n].
\] 
On the other hand, the continuum limit of the first term 
reads (see Appendix.\ref{Continuity-limit-underdamped})
\begin{align}
&\lim_{n\rightarrow +\infty}\sum_{i=0}^{n-1}\ln
\frac{\Pr[\tilde X_{i+1}=X_{i+1}|\tilde X_i=X_i]}
{\Pr[\tilde X_{n-i}^-=I(X_{i})|\tilde X_{n-i-1}^-=I(X_{i+1})]}\notag\\
&=-\frac{1}{T}\int_{t_I}^{t_F}\D t\frac{p^\mu}{m}\nabla_\mu\(\mathcal{T}+U\) 
+\frac{1}{T}\int_{t_I}^{t_F}\D t\frac{p_\mu}{m}f^\mu_{\text{noc}},
\label{eq34}
\end{align}
where $\mathcal{T}:=p^\mu p_\mu /2m$ is the kinematic energy.
According to eq.~\eqref{underdamped-heat-work}, the trajectory work should be 
\begin{align*}
\Delta W_{X_{[t]}}=\int_{t_I}^{t_F}\D t 
\(\frac{p_\mu}{m}f^\mu_{\text{noc}}+\pfrac{U}{\lambda}\frac{\D\lambda}{\D t} \),
\end{align*}
and the change of energy is
\begin{align*}
\Delta E_{X_{[t]}}&=\Delta\mathcal{T}_{X_{[t]}}+\Delta U_{X_{[t]}}\nonumber\\
&=\int_{t_I}^{t_F}\D t\frac{p^\mu}{m}\nabla_\mu \mathcal{T}
+\int_{t_I}^{t_F}\D t\(\frac{p^\mu}{m}\nabla_\mu U 
+\pfrac{U}{\lambda}\frac{\D\lambda}{\D t} \).
\end{align*}
Since the trajectory heat is $\Delta Q_{X_{[t]}}=\Delta E_{X_{[t]}}-\Delta W_{X_{[t]}}$, 
eq.~\eqref{SigmaXt} can also be rewritten as 
\begin{align}\label{fluctuation-underdamped}
\Sigma_{X_{[t]}}=\Delta S_{X_{[t]}}-\frac{1}{T}\Delta Q_{X_{[t]}}
=\Delta S_{X_{[t]}} + \Delta S_{\rm Res},
\end{align}
where, \blue{since the heat reservoir maintains in equilibrium, the Clausius equality holds, 
the change of the entropy of the heat reservoir reads}
\[
\Delta S_{\rm Res}=\Delta Q_{\rm Res}/T = -\Delta Q_{X_{[t]}}/T.
\] 
Thus $\Sigma_{X_{[t]}}$ is indeed the total entropy production.
Inserting eq.~\eqref{fluctuation-underdamped} back into eq.~\eqref{Pfb}, 
we get the desired FT
\begin{align}
\frac{\Pr[\tilde X_{[t]}= X_{[t]}]}{\Pr[\tilde X^-_{[t]}=X^-_{[t]}]}
=\re^{\Sigma_{X_{[t]}}}=\re^{\Delta S_{X_{[t]}}-\Delta Q_{X_{[t]}}/T},
\label{fluct1}
\end{align}
which tells that the process with positive total entropy production 
is probabilistically more preferred.

Taking the trajectory average of eq.~\eqref{fluct1},  
we get, by use of the Jensen inequality, the following result,
\begin{align}\label{fluctuation-intergral-underdamped}
\re^{-\mean{\Sigma_{\tilde X_{[t]}}}}&\leq\mean{\re^{-\Sigma_{\tilde X_{[t]}}}}
=\int\mathcal{D}[X_{[t]}]\Pr[\tilde X^-_{[t]}=X^-_{[t]}]\nonumber\\
&=\int\mathcal{D}[X^-_{[t]}]\Pr[\tilde X^-_{[t]}=X^-_{[t]}]=1,
\end{align}
where $\mathcal{D}[X_{[t]}]=\D X_0\wedge\D X_1\wedge...\wedge\D X_n$ is 
the measure on the trajectory space. It is easy to prove that 
the map $X_{[t]}\mapsto X^-_{[t]}$ preserves the measure, \ie 
$\mathcal{D}[X_{[t]}]=\mathcal{D}[X^-_{[t]}]$. 
Eq.\eqref{fluctuation-intergral-underdamped} is the so-called integral 
FT, which tells that the entropy production is 
non-negative in any macroscopic process, \ie $\mean{\Sigma_{\tilde X_{[t]}}}\geq 0$. 

The FT in the overdamped case can be constructed 
following a similar fashion, however the processes must be described solely
in configuration space. The definitions of the reversed process $\tilde x^-_{[t]}$ and 
the reversed trajectory $x^-_{[t]}$ are similar to the underdamped case, 
with the replacement $X_i\to x_i$. 
Therefore, 
\begin{align*}
&\Sigma_{x_{[t]}}\blue{:=}
\ln\frac{\Pr[\tilde x_{[t]}=x_{[t]}]}{\Pr[\tilde x^-_{[t]}=x^-_{[t]}]}
\nonumber\\
&=\sum_{i=0}^{n-1}\ln\frac{\Pr[\tilde x_{i+1}=x_{i+1}|\tilde x_i=x_i]}
{\Pr[\tilde x_{n-i}^-=x_{i}|\tilde x_{n-i-1}^-=x_{i+1}]}
+\ln\frac{\Pr[\tilde x_0= x_0]}{\Pr[ \tilde x_n= x_n]}.
\end{align*}
In the continuum limit, we have (see Appendix.\ref{Continuity-limit-overdamped})
\begin{align}
&\lim_{n\rightarrow +\infty}\sum_{i=0}^{n-1}
\ln\frac{\Pr[\tilde x_{i+1}=x_{i+1}|\tilde x_i=x_i]}
{\Pr[\tilde x_{n-i}^-=x_{i}|\tilde x_{n-i-1}^-=x_{i+1}]}\nonumber\\
&\qquad =\frac{1}{T}\int_{t_I}^{t_F} v_\mu f^\mu_{\text{ex}}\D t,
\label{eq41}
\end{align}
where $v^\mu$ is the velocity of the Brownian particle. 
\red{According to eq.~\eqref{trajectory-heat}}, 
the trajectory heat is
\begin{align*}
&\Delta Q_{x_{[t]}}=-\int_{t_I}^{t_F}v_\mu f^\mu_{\text{ex}}\D t,
\end{align*}
and the trajectory entropy production is 
\begin{align*}
&\Delta S_{x_{[t]}}=\ln\rho_{t_I}(x_0)-\ln\rho_{t_F}(x_n)
=\ln\frac{\Pr[\tilde x_0= x_0]}{\Pr[ \tilde x_n= x_n]}.
\end{align*}
Finally, we arrive at the desired FT
\begin{align*}
\frac{\Pr[\tilde x_{[t]}=x_{[t]}]}{\Pr[\tilde x^-_{[t]}=x^-_{[t]}]}
=\re^{\Sigma_{x_{[t]}}}=\re^{\Delta S_{x_{[t]}}-\Delta Q_{x_{[t]}}/T}.
\end{align*}
The integral FT in the overdamped case can be obtained in complete analogy
to the underdamped case, therefore, there is no need to repeat the construction.

\section{Concluding remarks}
\label{Conclusion}

The covariant LE and FPE on a Riemannian manifold are constructed
in both underdamped and overdamped cases. The concepts of trajectory heat and 
trajectory work are clarified, and the first law on the trajectory 
level is established. The Stratonovich coupling plays an important role in 
establishing the appropriate form of the first law. In either cases, the corresponding FTs 
are proved in both differential and integral forms. These results allow for a 
complete extension of the existing stochastic thermodynamics to arbitrary 
Riemannian manifolds, which in turn may be helpful in understanding the origin of
irreversibility in certain biological scenarios. 

During the construction, we also clarified the link between the 
PDF with the ensemble and trajectory entropy productions, which may also 
shed some light in the parallel constructions in flat spaces. Moreover, 
the different forms of the additional stochastic forces in the 
underdamped and overdamped cases are also worth of notice. Last, we hope the results 
presented here could be inspiring for an ultimate resolution for the fully 
general relativistic construction for the FTs.

\section*{Acknowledgement}

This work is supported by the National Natural Science Foundation of China under the grant
No. 12275138.

\providecommand{\href}[2]{#2}\begingroup
\footnotesize\itemsep=0pt
\providecommand{\epr}[2][]{\href{http://arxiv.org/abs/#2}{arXiv:#2}}


\appendix

\onecolumngrid

\section{Overdamped Fokker-Planck equation}
\label{overdamped_limit}
Here we outline the procedure for taking the overdamped limit following the line of 
Ref.~\cite{durang2015overdamped}. 
Using the diffusion operator 
\[
 A=\frac{\delta^{ab}}{2}L_a L_b+L_0\quad\text{with }L_0
 =\frac{1}{m}\mathcal{L}-\frac{1}{m}K^{\mu}{}_{\nu}p^\nu\frac{\partial}{\partial p^\mu}, 
 \quad L_a=R^{\mu}{}_{a}\frac{\partial}{\partial p^\mu},
\] 
the underdamped FPE can be written as 
\begin{align}\label{bar_fokkerplank}
\partial_t\Phi_t=A^\dagger\Phi_t,
\end{align}
where $A^\dagger$ is the adjoint of A. The FPE can be rewritten as
\begin{align*}
\partial_t\bar{\Phi}_t=\bar{A}^\dagger\bar{\Phi}_t,
\end{align*}
where $\bar \Phi_t:=[P^s]^{-{1}/{2}}\Phi_t$ and 
$\bar{A}^\dagger:=[P^s]^{-{1}/{2}}A^\dagger [P^s]^{{1}/{2}}$, 
and $P^s$ is given in eq.~\eqref{local-equilibrium}. 
$\bar{A}^\dagger$ can be decomposed in terms of the 
creation and annihilation operators
\begin{align*}
&a_\mu=\zeta\pfrac{}{p^\mu}+\frac{1}{2\zeta}p_\mu,\qquad
a^\dagger_\mu=-\zeta\pfrac{}{p^\mu}+\frac{1}{2\zeta}p_\mu
\end{align*}
which obey the commutation relation $[a_\mu,a_\nu^\dagger]=g_{\mu\nu}$, 
wherein $\zeta=\sqrt{mT}$.

Since $K^{\mu\nu}$ is a symmetric tensor, its eigenvectors 
$e\ind{\mu}{\hat{\nu}}$ constitute an orthonormal basis. 
The components of a tensor under the orthonormal basis are 
denoted by adding a hat on its index, \eg 
$W_{\hat{\mu}}=e\ind{\nu}{\hat{\mu}}W_\nu$,
$V_{\hat{\mu}}=e_{\nu\hat{\mu}} V^\nu$, 
where $e_{\nu\hat\mu}$ is the dual basis. For convenience, only lower indices are used 
under the orthonormal basis. The commutator between the creation and the 
annihilation operators can be rewritten as 
$[a_{\hat{\mu}},a^\dagger_{\hat{\nu}}]=\delta_{\hat{\mu}\hat{\nu}}$.
Moreover, $p_\mu$ and $\pfrac{}{p^\mu}$ can be decomposed as
\begin{align*}
p_\mu=\zeta e_{\mu\hat{\nu}}(a_{\hat{\nu}}+a^\dagger_{\hat{\nu}}),\qquad 
\pfrac{}{p^\mu}=\frac{1}{2\zeta}e_{\mu\hat{\nu}}
(a_{\hat{\nu}}-a^\dagger_{\hat{\nu}}).
\end{align*} 
Using the above notations, one has
\begin{align*}
\bar{A}^\dagger=-\frac{1}{m}\kappa^{\hat{\mu}}N_{\hat{\mu}}
+\frac{1}{\zeta}f^{\hat{\mu}}_{\text{ex}}a^\dagger_{\hat{\mu}}
-\frac{1}{m}\mathcal{L},
\end{align*}
where $\kappa^{\hat\mu}$ represent the eigenvalues of $K^{\mu\nu}$, and 
$N_{\hat \mu}:=a^\dagger_{\hat\mu}a_{\hat\mu}$. Let 
$\psi_0:=[P^s]^{{1}/{2}}$, $\psi_{\hat\mu}:=a^\dagger_{\hat\mu}\psi_0$. 
Clearly, we have $a_{\hat\mu}\psi_0=0$, \ie the ground state $\psi_0$  
of the Fock space generated by $a^\dagger_{\hat\mu}$ 
corresponds to the equilibrium distribution $P^s$, and 
the excited states to non-equilibrium modifications.

In principle,  $\bar\Phi_t $ can be expanded as a linear superposition of 
the eigenstates of $N_{\hat \mu}$. However, in the overdamped limit, 
the time resolution $\Delta t$ is considered to be much larger than the 
relaxation time, so, in the first-order approximation near 
equilibrium state, we have
\begin{align}
\bar{\Phi}_t(x,p)\approx C^0(x,t)\psi_0(x,p)+C^{\hat\mu}(x,t)\psi_{\hat\mu}(x,p).
\label{formofsolution}
\end{align}

Some important commutation relations are listed below:
\begin{align}
&[p^\alpha,a^\dagger_{\hat\mu}]=\zeta e\ind{\alpha}{\hat\mu},\quad
[\pfrac{}{p^\alpha},a^\dagger_{\hat\mu}]=\frac{1}{2\zeta}e_{\alpha\hat\mu},\nonumber\\
&[\pfrac{}{x^\alpha},a^\dagger_{\hat\mu}]=\frac{1}{2\zeta}
\partial_\alpha g_{\nu\beta}p^\beta e\ind{\nu}{\hat\mu}
+a^\dagger_\nu\partial_\alpha e\ind{\nu}{\hat\mu},\nonumber\\
&[\mathcal{L},a^\dagger_{\hat\mu}]=\zeta e\ind{\sigma}{\hat\mu}
\left[\pfrac{}{x^\sigma}-\Gamma\ind{\nu}{\alpha\sigma}p^\alpha\pfrac{}{p^\nu} \right]
+\zeta e_{\nu\hat\alpha}\nabla_{\hat\beta} e\ind{\nu}{\hat\mu}(a_{\hat\beta}
+a^\dagger_{\hat\beta})a^\dagger_{\hat\alpha}.
\label{com_Liouville}
\end{align}
Using eq.~\eqref{com_Liouville} one gets 
\begin{align*}
\mathcal{L}(\bar\Phi_t)&=\mathcal{L}(C^0)\psi_0+\mathcal{L}(C^{\hat\mu})\psi_{\hat\mu}
+C^{\hat\mu} \mathcal{L}(\psi_{\hat\mu})
=\mathcal{L}(C^0)\psi_0+\mathcal{L}(C^{\hat\mu})\psi_{\hat\mu}
+C^{\hat\mu}[\mathcal{L},a^\dagger_{\hat\mu}]\psi_0\notag\\
&=\mathcal{L}(C^0)\psi_0+\mathcal{L}(C^{\hat\mu})\psi_{\hat\mu}
+\zeta\nabla_\nu e\ind{\nu}{\hat\mu}C^{\hat\mu}\psi_0
+\zeta e_{\nu\hat\alpha}(\nabla_{\hat\beta}e\ind{\nu}{\hat\mu})C^{\hat\mu}
(a^\dagger_{\hat\beta}a^\dagger_{\hat\alpha}\psi_0),
\end{align*}
where the property $\(\pfrac{}{x^\sigma}-\Gamma\ind{\nu}{\alpha\sigma}
p^\alpha\pfrac{}{p^\nu} \)\psi^0=0$ has been used. 
Since $\mathcal{L}(C^0)$ and $\mathcal{L}(C^{\hat\mu})$ still contain momentum, 
these expressions can be further expanded,
\begin{align*}
&\mathcal{L}(C^0)\psi_0
=\partial^{\hat\mu} C^0 p_{\hat\mu}\psi_0
=\zeta \nabla^{\hat\mu} C^0 \psi_{\hat\mu},\\
&\mathcal{L}(C^{\hat\nu})\psi_{\hat\nu}=\zeta \nabla_{\hat\mu} C^{\hat\mu} \psi_{0}
+\zeta \nabla_{\hat\alpha} C^{\hat\beta} 
(a^\dagger_{\hat\alpha}a^\dagger_{\hat\beta}\psi_0).
\end{align*}
Defining an operator $D_{\hat\mu}:=(\zeta/T)(f^{\hat\mu}_{\text{ex}}
-T\partial_{\hat\mu})$ and substituting eq.~\eqref{formofsolution} into 
eq.~\eqref{bar_fokkerplank}, the evolution equations of $C^0$ and $C^{\hat\mu}$ 
follow,
\begin{align}
\partial_t C^0&=-\frac{T}{\zeta}\left[\nabla_{\hat\mu}C^{\hat\mu}
+(\nabla_\nu e\ind{\nu}{\hat\mu})C^{\hat\mu} \right],\label{73}\\
\partial_t C^{\hat\mu}&=-\frac{\kappa^{\hat\mu}}{m}C^{\hat\mu}
+\frac{1}{m}D_{\hat\mu}C^0.\label{74}
\end{align}
The overdamped limit means that ${\kappa}^{\hat \mu}$ is large. In such a limit, 
$C^0$ and $C^{\hat\mu}$ are respectively of orders $O(({\kappa}^{\hat \mu})^0)$ and
$O(({\kappa}^{\hat \mu})^{-1})$. Therefore, up to order $O(({\kappa}^{\hat \mu})^0)$,
we can safely ignore the left hand side of eq.~\eqref{74}, yielding 
\[
C^{\hat\mu} \approx \frac{1}{\kappa^{\hat\mu}} D_{\hat\mu} C^0.
\]
Inserting this result into eq.~\eqref{73}, we get
\begin{align}
\partial_t C^0&= -\frac{T}{\zeta}
\left[\nabla_{\hat\mu}\(\frac{1}{\kappa^{\hat\mu}}D_{\hat\mu}C^{0}\)
+(\nabla_\nu e\ind{\nu}{\hat\mu})\frac{1}{\kappa^{\hat\mu}} D_{\hat\mu}C^{0}\right]
= -\frac{T}{\zeta}\nabla_{\nu}\(e\ind{\nu}{\hat\mu}\frac{1}{\kappa^{\hat\mu}}
e\ind{\alpha}{\hat\mu}D_{\alpha}C^{0}\)\notag\\
    &= -\nabla_\nu\hat K^{\nu\alpha}\left[g_{\alpha\mu}f^\mu_{\text{ex}}C^0-T\nabla_\alpha C^0 \right] \notag\\
    &=\nabla_\mu\left[\frac{1}{2}\hat D^{\mu\nu}\nabla_\nu C^0-\hat K\ind{\mu}{\nu}f^{\nu}_{\text{ex}}C^0 \right],
\label{eqC0}
\end{align} 
where $\hat{K}^{\mu\nu}=e\ind{\mu}{\hat\alpha}e\ind{\nu}{\hat\alpha}
(\kappa^{\hat\alpha})^{-1}$ is inverse of the damping tensor $K^{\mu\nu}$, 
and $\hat D^{\mu\nu}:=2T\hat{K}^{\mu\nu}=4T^2(D^{-1})^{\mu\nu}$ is the 
rescaled diffusion tensor. Since 
\begin{align*}
&\rho_t(x)=\int g^{{1}/{2}} \D^d p~\Phi_t(x,p)
=\int g^{{1}/{2}} \D^d p~\psi_0(x,p)\bar\Phi_t(x,p)=C^0(x,t),
\end{align*}
the overdamped FPE \eqref{fokkerplank2} follows from eq.~\eqref{eqC0}.

\section{Continuum limit}
\label{Continuity-limit}

Let us first introduce some mathematical tricks. 
Let $A$ be a full-rank square matrix and $B$ be a matrix of the same size, 
the determinant of $A+B\D t$ can be expanded as
\begin{align}\label{58}
\det[A+B\D t+O(\D t^2)]&=\det[A]\det[I+A^{-1}B\D t+O(\D t^2)]
=\det[A]+\det[A]\Tr[A^{-1}B]\D t+O(\D t^2).
\end{align}
Let $f(t)$ be a continuous function on $[t_I,t_F]$. Then 
there is a continuum limit
\begin{align}\label{59}
\lim_{n\rightarrow +\infty}\prod_{i=0}^{n-1}[1+f(t_i)\D t+O(\D t^2)]
=\exp\left[\int_{t_I}^{t_F}\D tf(t) \right],
\end{align}
where $\D t=(t_F-t_I)/n$ and $t_i\in [i\D t,(i+1)\D t]$. 
Combining eqs.~\eqref{58}-\eqref{59}, we have
\begin{align}\label{93}
&\lim_{n\rightarrow +\infty}\prod_{i=0}^{n-1}\frac{\det[A(t_i)+B(t_i)\D t+O(\D t^2)]}
{\det[A(t_i)+C(t_i)\D t+O(\D t^2)]}
=\exp\left[\int_{t_I}^{t_F}\D t\Tr[A^{-1}(B-C)] \right],
\end{align}
where $A(t),B(t),C(t)$ are continuous matrix functions on $[t_I,t_F]$.

\subsection{Overdamped case}
\label{Continuity-limit-overdamped}

The discrete time version of the OLE 
\eqref{la-overdamped-continuity} reads
\begin{align*}
\tilde x^\mu_{i+1}-\tilde x^\mu_i
=\hat{F}^\mu(\bar\lambda_i)|_{(\tilde x_{i+1}+\tilde x_i)/2}\D t
+\hat R\ind{\mu}{a}|_{(\tilde x_{i+1}+\tilde x_i)/2}\D\tilde w^a_i,
\end{align*}
where $\bar \lambda_i:=(\lambda_{i+1}+\lambda_i)/2$. 
The choice of discretization rule for $\hat F^\mu$ does not affect the continuum limit, 
but we still take the middle-point rule for consistency of notations. 
We introduce a function
\begin{align*}
\D w^a(x,y,\lambda):=(\hat R^{-1})\ind{a}{\nu}|_{\bar{x}}\left[x^\nu-y^\nu
+\hat F^\nu(\lambda)|_{\bar x}\D t  \right],
\end{align*}
where $\bar x:=(x+y)/2$. This function is connected with 
the Gaussian noise $\D\tilde w^a_i$ via the relation
\begin{align}\label{96}
\D \tilde w_i^a=\D w^a(\tilde x_{i+1},\tilde x_{i},\bar \lambda_i).
\end{align}
The reversed process is governed by same LE with the external 
control parameter time-reversed, so that
\begin{align*}
\D \tilde w^a_i=\D w^a(\tilde x^-_{i+1},\tilde x^-_i,\bar \lambda_i^-)
=\D w^a(\tilde x^-_{i+1},\tilde x^-_i,\bar \lambda_{n-i-1}).
\end{align*}
The last equation can also be rewritten as 
\begin{align}\label{98}
\D \tilde w^a_{n-i-1}=\D w^a(\tilde x^-_{n-i},\tilde x^-_{n-i-1},\bar \lambda_{i}).
\end{align}
Using eqs.~\eqref{96} and \eqref{98}, the transition probabilities of 
the forward and reversed processes can be expressed as
\begin{align}\label{99}
&\Pr[\tilde x_{i+1}=x_{i+1}|\tilde x_i=x_i] 
= g^{-{1}/{2}}(x_{i+1})\det\left[\pfrac{\D w^a}{x^\mu} \right]
(x_{i+1},x_i,\bar\lambda_i)\Pr[\D\tilde w^a_i=\D w^a(x_{i+1},x_i,\bar\lambda_i)]\\
\label{100}
&\Pr[\tilde x_{n-i}^-=x_{i}|\tilde x_{n-i-1}^-=x_{i+1}]
= g^{-{1}/{2}}(x_{i})\det\left[\pfrac{\D w^a}{x^\mu} \right]
(x_i,x_{i+1},\bar\lambda_i)
\Pr[\D\tilde w^a_{n-i-1}=\D w^a(x_i,x_{i+1},\bar\lambda_i)].
\end{align}
It is easy to calculate the Jacobi matrix of $\D w^a$,
\begin{align*}
\pfrac{\D w^a}{x^\mu}(x_{i+1},x_i,\bar\lambda_i)
&=(\hat R^{-1})\ind{a}{\mu}|_{\bar x_i} 
+\frac{1}{2}\left\{\partial_\mu(\hat R^{-1})\ind{a}{\nu} v^\nu_i
-\partial_\mu[(\hat R^{-1})\ind{a}{\nu}\hat F^\nu(\bar \lambda_i)] 
\right\}_{\bar{x}_i}\D t,\\
\pfrac{\D w^a}{x^\mu}(x_i,x_{i+1},\bar\lambda_i)
&=(\hat R^{-1})\ind{a}{\mu}|_{\bar x_i} 
+\frac{1}{2}\left\{-\partial_\mu(\hat R^{-1})\ind{a}{\nu} v^\nu_i
-\partial_\mu[(\hat R^{-1})\ind{a}{\nu}\hat F^\nu(\bar\lambda_i)]  
\right\}_{\bar{x}_i}\D t,
\end{align*}
where $v_i:=(x_{i+1}-x_i)/\D t$ is the velocity of the Brownian particle. 
Using eq.~\eqref{93}, we can get
\begin{align}
\lim_{n\rightarrow +\infty}\prod_{i=0}^{n-1}
\frac{g^{-1/2}(x_{i+1})\det\left[\pfrac{\D w^a}{x^\mu} \right]
(x_{i+1},x_i,\bar\lambda_i)}
{g^{-1/2}(x_i)\det\left[\pfrac{\D w^a}{x^\mu} \right]
(x_i,x_{i+1},\bar\lambda_i)}
=\exp\left[ -\int_{t_I}^{t_F}v^\nu(\hat R^{-1})\ind{a}{\nu}\nabla_\mu
\hat R\ind{\mu}{a}\D t\right].
\end{align}
Therefore, using eq.~\eqref{wienner}, we arrive at the following continuum limit,
\begin{align}\label{106}
&\lim_{n\rightarrow +\infty}\prod_{i=0}^{n-1}
\frac{\Pr[\D\tilde w^a_i=\D w^a(x_{i+1},x_i,\bar\lambda_i)]}
{\Pr[\D\tilde w^a_{n-i-1}=\D w^a(x_i,x_{i+1},\bar\lambda_i)]}
=\exp\left[\frac{1}{T}\int_{t_I}^{t_F}v_\mu f^\mu_{\text{ex}}\D t \right]
\exp\left[\int_{t_I}^{t_F}v^\nu (\hat R^{-1})\ind{a}{\nu}
\nabla_\mu\hat R\ind{\mu}{a}\D t \right].
\end{align}
Finally, using eqs.~\eqref{99}-\eqref{100} and \eqref{106}, we get
\begin{align*}
&\lim_{n\rightarrow +\infty}\prod_{i=0}^{n-1}
\frac{\Pr[\tilde x_{i+1}=x_{i+1}|\tilde x_i=x_i]}
{\Pr[\tilde x_{n-i}^-=x_{i}|\tilde x_{n-i-1}^-=x_{i+1}]}\
=\exp\left[\frac{1}{T}\int_{t_I}^{t_F} v_\mu f^\mu_{\text{ex}}\D t \right],
\end{align*}
which is essentially identical to eq.~\eqref{eq41}.

\subsection{Underdamped case}
\label{Continuity-limit-underdamped}

Similarly, \red{we write down the discrete time version of 
eqs.~\eqref{la3.1} and \eqref{la3.3}},
\begin{align*}
\tilde x^\mu_{i+1}-\tilde x^\mu_i&=\frac{\tilde p^\mu_{i+1}+\tilde p^\mu_i}{2m}\D t,\\
\tilde p^\mu_{i+1}-\tilde p^\mu_i
&=F^\mu(\bar \lambda_i)|_{(\tilde X_{i+1}+\tilde X_i)/2}\D t 
+\bar F^\mu|_{(\tilde X_{i+1}+\tilde X_i)/2}\D t
+R\ind{\mu}{a}|_{(\tilde X_{i+1}+\tilde X_i)/2}\D\tilde w^a_i.
\end{align*}
Now $\tilde x_{i+1}$ should be viewed as a function in $\tilde p_{i+1}$ and $\tilde X_i$, 
\begin{align*}
\tilde x_{i+1}=x(\tilde p_{i+1},\tilde X_i)
=\frac{\tilde p_{i+1}+\tilde p_i}{2m}\D t+\tilde x_i.
\end{align*}
Denoting $\Delta(X_{i+1},X_i)=\delta^d(x_{i+1}-x_i-(p_{i+1}+p_i)\D t/2m)$, 
we have the conditional probability
\begin{align*}
&\Pr[\tilde x_{i+1}=x_{i+1}|\tilde p_{i+1}=p_{i+1},\tilde X_i=X_i]
=g^{-1/2}(x_{i+1})\Delta(X_{i+1},X_i).
\end{align*}
Notice that $\Delta(X_{i+1},X_i)=\Delta(I(X_i),I(X_{i+1}))$. Defining 
\begin{align*}
&\D w^a(X,Y,\lambda):=(R^{-1})\ind{a}{\nu}|_{\bar X}
\left[p^\nu-k^\nu-F^\nu(\lambda)|_{\bar X}\D t+\bar F^\nu|_{\bar X}\D t \right],
\end{align*}
where $X=(x,p)$, $Y=(y,k)$ and $\bar X=(X+Y)/2$, we have
\begin{align*}
\D \tilde w^a_i=\D w^a(\tilde X_{i+1},\tilde X_i,\bar\lambda_i)
=\D w^a(\tilde p_{i+1},x(\tilde p_{i+1},\tilde X_i),\tilde X_i,\bar\lambda_i).
\end{align*}
Similarly, for the reversed process, we have 
\begin{align*}
\D\tilde w^a_{n-i-1}=\D w^a(\tilde p_{n-i}^-,x(\tilde p_{n-i}^-,\tilde X_{n-i-1}^-),
\tilde X^-_{n-i-1},\bar\lambda_i).
\end{align*}
Now introduce the matrix
\begin{align*}
T\ind{a}{\mu}(X,Y,\lambda)
&:=\pfrac{\D w^a}{p^\mu}(X,Y)+\pfrac{x^\nu}{p^\mu}\pfrac{\D w^a}{x^\nu}(X,Y)\notag\\
&=\pfrac{\D w^a}{p^\mu}(X,Y)+\frac{1}{2m}\pfrac{\D w^a}{x^\mu}(X,Y)\D t.
\end{align*}
The transition probabilities of the forward and reversed processes can be written as 
\begin{align}\label{117}
&\Pr[\tilde X_{i+1}=X_{i+1}|\tilde X_i=X_i]
=\Pr[\tilde x_{i+1}=x_{i+1}|\tilde p_{i+1}=p_{i+1},\tilde X_i=X_i]
\Pr[\tilde p_{i+1}=p_{i+1}|\tilde X_{i}=X_i],\notag\\
&\qquad= g^{-1}(x_{i+1})\Delta(X_{i+1},X_i)\det[T\ind{a}{\mu}]
(X_{i+1},X_i,\bar\lambda_i)
\Pr[\D\tilde w_i^a=\D w^a(X_{i+1},X_i,\bar\lambda_i)],
\end{align}
and 
\begin{align}\label{118}
&\Pr[\tilde X_{n-i}^-=I(X_i)|\tilde X_{n-i-1}^-=I(X_{i+1})]
=g^{-1}(x_i)\Delta(I(X_i),I(X_{i+1}))\det[T\ind{a}{\mu}]
(I(X_i),I(X_{i+1}),\bar\lambda_i)\notag\\
&\qquad \times \Pr[\D\tilde w_i^a=\D w^a(I(X_{i}),I(X_{i+1}),\bar\lambda_i)].
\end{align}
The Jacobi matrices in $T\ind{a}{\mu}(X,Y,\lambda)$ can be explicitly calculated,
yielding 
\begin{align*}
&T\ind{a}{\mu}(X_{i+1},X_i,\bar\lambda_i)
=(R^{-1})\ind{a}{\mu}|_{\bar X_i}
+\frac{1}{2}\left\{m\pfrac{}{p^\mu}(R^{-1})\ind{a}{\nu} A^\nu_i
-\pfrac{}{p^\mu}[(R^{-1})\ind{a}{\nu}F^\nu(\bar\lambda_i)] 
-\pfrac{}{p^\mu}[(R^{-1})\ind{a}{\nu}\bar F^\nu]  
\right\}_{\bar X_i}\D t,\\
&T\ind{a}{\mu}(I(X_{i}),I(X_{i+1}),\bar\lambda_i)
=(R^{-1})\ind{a}{\mu}|_{\bar X_i} 
+\frac{1}{2}\left\{-m\pfrac{}{p^\mu}(R^{-1})\ind{a}{\nu} A^\nu_i
+\pfrac{}{p^\mu}[(R^{-1})\ind{a}{\nu}F^\nu(\bar\lambda_i)] 
-\pfrac{}{p^\mu}[(R^{-1})\ind{a}{\nu}\bar F^\nu]  
\right\}_{\bar X_i}\D t,
\end{align*}
where $A^\mu_i:=(p^\mu_{i+1}-p_i^\mu)/(m\D t)$ is the coordinates acceleration, 
and terms of order $O(\D t^2)$ are omitted. Using these results together with 
eq.~\eqref{93}, we can get
\begin{align*}
&\lim_{n\rightarrow +\infty}\prod_{i=0}^{n-1}\frac{\det[T\ind{a}{\mu}]
(X_{i+1},X_i,\bar\lambda_i)}{\det[T\ind{a}{\mu}](I(X_{i}),I(X_{i+1}),\bar\lambda_i)}
\frac{g^{-1}(x_{i+1})}{g^{-1}(x_i)}\notag\\
&\qquad =\exp \left[ \int^{t_F}_{t_I}\(R\ind{\mu}{a}\pfrac{}{p^\mu}
[(R^{-1})\ind{a}{\nu}(ma^\nu-f^\nu_{\text{ex}})]
+\frac{2}{m}\Gamma\ind{\mu}{\mu\alpha}p^\alpha \) \D t \right]
\exp\left[-\int^{t_F}_{t_I}g^{-1}\frac{p^\mu}{m}\pfrac{}{x^\mu}g\D t \right]\notag\\
&\qquad = \exp \left[ \int^{t_F}_{t_I}R\ind{\mu}{a}
\pfrac{}{p^\mu}(R^{-1})\ind{a}{\nu}(ma^\nu-f^\nu_{\text{ex}}) \D t \right],
\end{align*}
where $a^\mu=A^\mu+\Gamma\ind{\mu}{\alpha\beta}p^\alpha p^\beta/m$ is 
covariant acceleration. Using eq.\eqref{wienner}, the following continuum limit 
can be derived,
\begin{align}\label{123}
&\lim_{n\rightarrow +\infty}\prod_{i=0}^{n-1}
\frac{\Pr[\D\tilde w_i^a=\D w^a(X_{i+1},X_i,\bar\lambda_i)]}
{\Pr[\D\tilde w_i^a=\D w^a(I(X_{i}),I(X_{i+1}),\bar\lambda_i)]}\notag\\
&\qquad = \exp\left[-\frac{1}{T}\int^{t_F}_{t_I}\frac{p_\mu}{m}
(m a^\mu-f^\mu_{\text{ex}})\D t \right]
\exp\left[-\int^{t_F}_{t_I}R\ind{\mu}{a}\pfrac{}{p^\mu}
(R^{-1})\ind{a}{\nu}(ma^\nu-f^\nu_{\text{ex}})\right].
\end{align}
Finally, using eqs.~\eqref{117} and \eqref{118}, we get
\begin{align*}
&\lim_{n\rightarrow +\infty}\prod_{i=0}^{n-1}
\frac{\Pr[\tilde X_{i+1}=X_{i+1}|\tilde X_i=X_i]}
{\Pr[\tilde X_{n-i}^-=I(X_{i})|\tilde X_{n-i-1}^-=I(X_{i+1})]}
= \exp\left[-\frac{1}{T}\int^{t_F}_{t_I}
\frac{p_\mu}{m}(m a^\mu-f^\mu_{\text{ex}})\D t \right]\notag\\
&\quad =\exp\left[-\frac{1}{T}\int_{t_I}^{t_F}\D t\frac{p^\mu}{m}
\nabla_\mu\(\mathcal{T}+U\) +\frac{1}{T}\int_{t_I}^{t_F}\D t
\frac{p_\mu}{m}f^\mu_{\text{noc}} \right].
\end{align*}
This result is identical to eq.~\eqref{eq34}.

\end{document}